# Magnetic Anisotropy and Its Structural Origins in Ru-Substituted Manganite Films

Brajagopal Das,[1] Lena Wysocki,[2] Jörg Schöpf,[2] Lin Yang,[2] Amir Capua,[3] Paul H.M. van Loosdrecht,[2] Lior Kornblum[1,*]

[1]*Andrew & Erna Viterbi Department of Electrical & Computer Engineering, Technion Israel Institute of Technology, 3200003 Haifa, Israel*

[2]*University of Cologne, Institute of Physics II, 50937 Cologne, Germany*

[3]*Department of Applied Physics, The Hebrew University of Jerusalem, Jerusalem 91904, Israel*

[*] liork@ee.technion.ac.il

## Abstract

Controlling magnetic anisotropy (MA) is important in a variety of applications including magnetic memories, spintronic sensors, and skyrmion-based data distribution. The perovskite manganite family provides a fertile playground for complex, intricate, and potentially useful structure-magnetism relations. Here we report on the MA that emerges in 10% Ru substituted $La_{0.7}Sr_{0.3}MnO_3$ (Ru-LSMO) films for which strong perpendicular magnetization and anisotropic in-plane magnetization are found. These moderately compressively strained films possess a rich microstructure, consisting of coherently strained phase which evolves into a one dimensional (1D) periodically-modulated structure above a critical thickness. We illustrate how 10% Ru substitution plays a crucial role behind the observed MA, and how the structural distortion and 1D periodic structural modulation produce the anisotropic in-plane magnetization. We highlight the practical significance of the observed MA, which could pave the way towards the realization of cutting-edge oxide-based room temperature spintronic memory devices.

## 1. INTRODUCTION

Mixed-valence manganites such as $La_{0.7}Sr_{0.3}MnO_3$ (LSMO) have attracted significant attention owing to their colossal magnetoresistance (CMR), metal-insulator transitions, ferromagnetism, high spin polarization near the Fermi level, and high Curie temperature ($T_c$~370 ºC in bulk) [1–5]. This system provides a textbook example of the complexity of structure-property relationships in correlated oxide systems, where small structural details can result in dramatic changes in the macroscopic behavior. Various properties of LSMO thin films, such as coercivity, ferromagnetic domains, magnetoresistance, and magnetization switching are related to the magnetic anisotropy, which can be tuned by several parameters, such as the growth conditions, substrate surface engineering, atomic substitution, thickness, temperature, and applied magnetic field [6–10]. The microstructure plays crucial roles in the manifestation of magnetic anisotropy, and therefore epitaxial strain provides a direct route for tuning the magnetic properties [11]. In addition, the steps and terraces on vicinal substrates can break the 4-fold rotational symmetry and reduce it into 2-fold rotational symmetry, thereby affecting the growth mechanism and modifying the microstructure; this can lead to the manifestation of in-plane uniaxial magnetic anisotropy [8,10]. These structural details and the resulting magnetic anisotropy can have a crucial impact on various magnetic anisotropy-based devices. Furthermore, atomic substitution can play a significant role in epitaxial strain engineering, thereby affecting the functional properties of a film.

Konoto et al. demonstrated Ru substitution-induced magnetic anisotropy in 5% Ru-substituted manganite ($La_{0.6}Sr_{0.4}Mn_{0.95}Ru_{0.05}O_3$) film grown on an STO substrate [12]. More recently, Nakamura et al. reported non-trivial magnetic topologies emerge in Ru-LSMO films when the perpendicular magnetization is controlled via Ru substitution and substrate-induced compressive strain [7]. They reported strong perpendicular magnetic anisotropy in 10% Ru-LSMO, with significantly reduced anisotropy when the Ru substitution is reduced to 5%. From a practical perspective, tilted magnetic anisotropy (TMA) with a strong perpendicular magnetization is attractive for several emerging memory and spintronic technologies, such as those based on spin-orbit torque (SOT) [13,14]. In addition to TMA with a strong perpendicular component, anisotropic in-plane magnetization is also required for practical applications, such as deterministic perpendicular magnetization switching through SOT [13,15].

Here we report TMA with strong perpendicular magnetization and anisotropic in-plane magnetization in 10% Ru-LSMO films. By detailed microstructural analysis, we unveil the microstructural origin of the MA. We observe and analyze one-dimensional (1D) periodic structural modulation in the thick 10% Ru-LSMO film. The relations between strain, microstructure and magnetism are discussed, illustrating the role of Ru and strain-induced structural mechanisms behind the MA. This study was carried out at low temperatures (30 K) where the various mechanisms can be readily identified.

## II. EXPERIMENTAL DETAILS

Ru-LSMO films were epitaxially grown on LSAT (001) substrates (Crystec GmbH) using pulsed laser deposition (PLD). The substrates were held at 650 °C and the target was ablated using a KrF laser with a fluence of 2.4 $J \cdot cm^{-2}$ at a repetition rate of 3-5 Hz. The oxygen pressure was maintained at ~0.13 mbar during growth, and it was increased to 100 mbar after growth while the samples were cooled down at a rate of 10 °C/min.

Temperature and field-dependent magnetization measurements were performed using a superconducting quantum interference device (SQUID) magnetometer in a magnetic properties measurement system (MPMS3, Quantum Design). X-ray diffraction (XRD) measurements were performed at room temperature using a Rigaku SmartLab diffractometer with Cu $K_\alpha$ radiation ($\lambda = 1.54$ Å) and a 2-bounce incident monochromator.

## III. RESULTS AND DISCUSSION

### III.A. MAGNETIC ANISOTROPY

The temperature dependence of the magnetization (M-T curves, Figs. 1a, S1a) and magnetic field-dependent magnetization loops (M-H curves, Fig. 1b) along the main lattice directions show that a 48 nm 10% Ru-LSMO film has strong perpendicular magnetization as well as anisotropic in-plane magnetization. This suggests that the easy axis of magnetization is tilted from the surface normal, implying TMA is found. We note that the in-plane projection of the easy axis does not lie along the main in-plane lattice directions. From the M-H curves (Fig. 1b, inset), the ascending order of magneto-crystalline anisotropy energy (E) can be estimated as: $E_{[001]pc} < E_{[010]pc} < E_{[100]pc}$, where 'pc' stands for pseudocubic lattice coordinates. Henceforth, we use the notation for the in-plane ([100], [010]) and the perpendicular ([001]) pseudocubic directions.

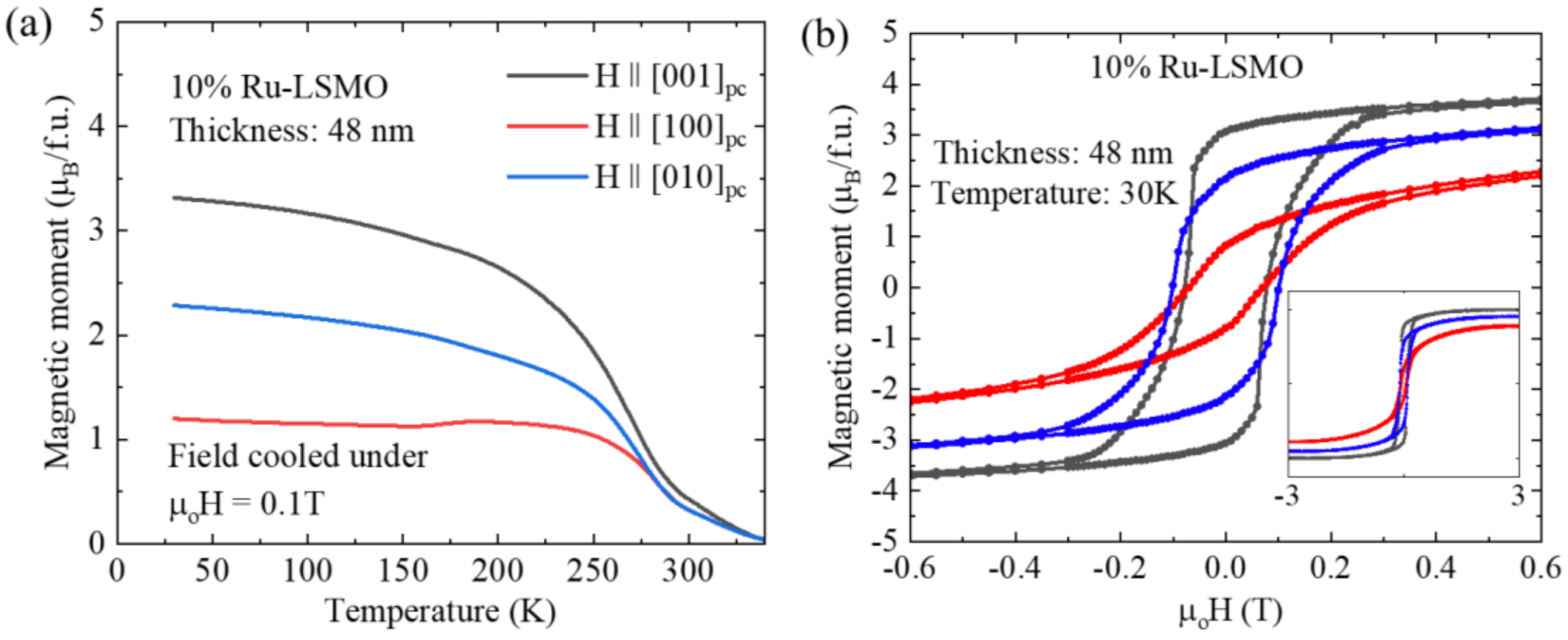

**Figure 1.** Magnetic properties of a 48 nm 10% Ru-LSMO film. (a) Magnetization as a function of temperature (M-T) curves along the main pseudocubic lattice directions. Before each M-T measurement, the sample was field cooled under 0.1 T and the measurement was performed during warm up under 0.1 T. (b) Magnetic field-dependent magnetization (M-H) loops at 30 K; before each M-H measurement, the sample was zero field cooled to 30 K. The inset shows these loops extended to ±3 T.

While TMA is also observed in a thin (10.5 nm) 10% Ru-LSMO film (Figs. S2, S1b), at this lower thickness the M-T and M-H behavior (Figs. S2, S1b) indicate that the magnetization anisotropy between the two main in-plane lattice directions is significantly diminished. In addition, we note the flattening of the $[100]_{pc}$ M-T curve of the thick 10% Ru-LSMO below ~200K, which includes a slight downturn (Fig. 1a) that is reproducible at higher magnetic fields (Fig. S1a). While further analysis is required to clarify the origins of this small feature, we observe it only in the sample featuring a 1D periodic structural modulation (to be discussed later on), suggesting a possible connection. Furthermore, we rule out any significant contribution of shape anisotropy by comparing the $[001]_{pc}$ M-H curves of a thick (48 nm) and a thin (10.5 nm) 10% Ru-LSMO films from 0 T to 3 T (Fig. S4).

Altogether, we observe a TMA with strong perpendicular magnetization and anisotropic in-plane magnetization in both the thick and thin Ru-LSMO films; however, anisotropy of the in-plane component in the thin film is negligible compared to that of the thick film. To explain these observations, we first study the microstructure of these films in Section III.B, and then in Section III.C we combine the magnetism and microstructure to explain the origins of magnetic anisotropy.

## III.B. MICROSTRUCTURE

To understand the structural origins of the observed TMA, off-specular XRD reciprocal space maps (RSMs) of the thick (48 nm) 10% Ru-LSMO film were acquired at room temperature. The results (Fig. 2) suggest that the main Bragg peaks of the film and the corresponding Bragg peaks of the substrate have the same in-plane momentum transfer $Q_{||}$, confirming coherent growth of fully-strained Ru-LSMO films (see Fig. S5 for the 2θ-ω Bragg peaks). Bulk LSMO has a rhombohedral crystal structure with a pseudo-cubic lattice parameter of 3.875 Å [16], which is only slightly larger than the lattice parameter of 3.868 Å of the cubic LSAT substrate. The substitution of Mn by Ru increases the lattice parameter [17], resulting in increased compressive strain (to be discussed later).

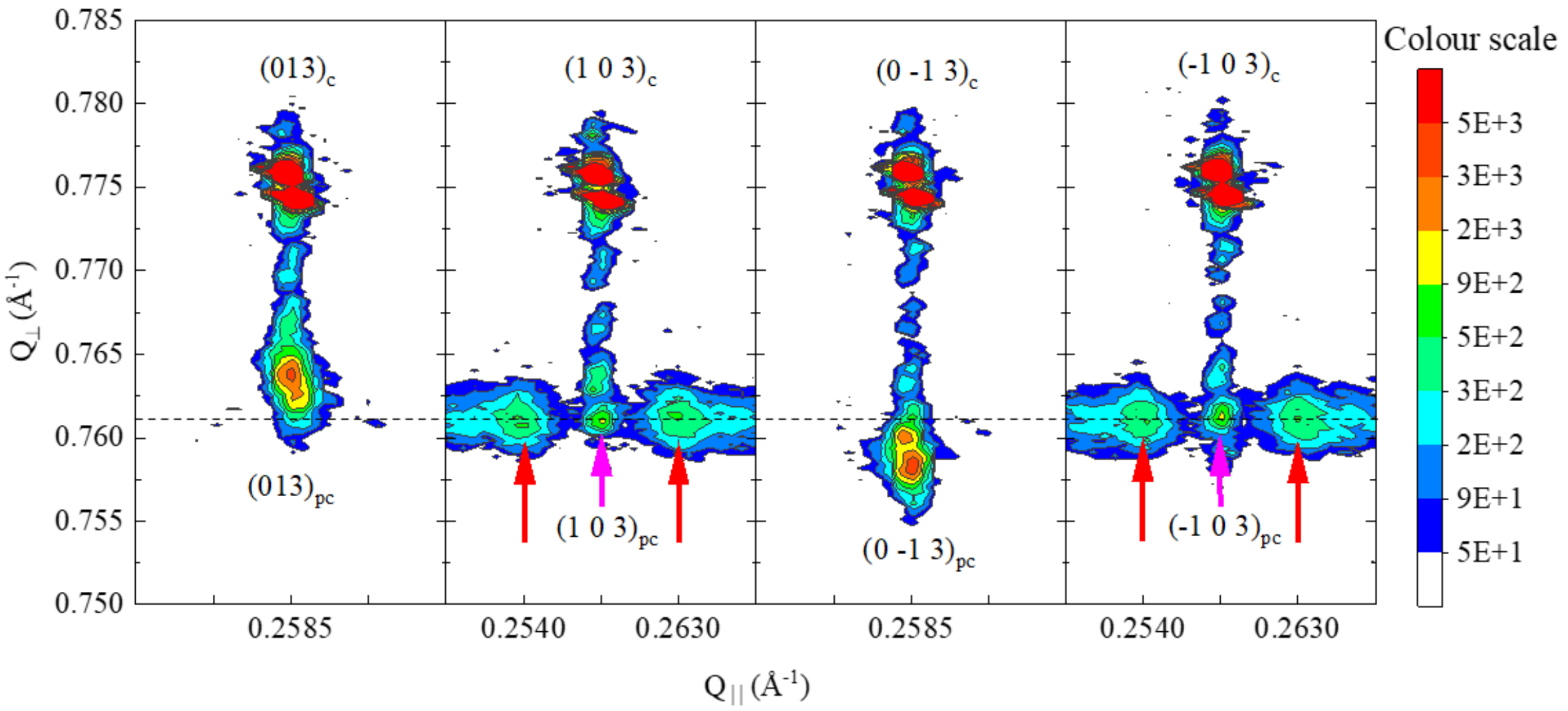

**Figure 2.** Off-specular RSMs of 48 nm 10% Ru-LSMO film around the $(0\ 1\ 3)_c$, $(1\ 0\ 3)_c$, $(0\ \text{-}1\ 3)_c$, and $(\text{-}1\ 0\ 3)_c$ reflections of LSAT(001). Intensities are presented on logarithmic scale; pink arrows indicate the film's main Bragg peaks, whereas red arrows indicate the corresponding satellites. The horizontal dashed line denotes the $Q_\perp$ position of the $(\pm 1\ 0\ 3)_{pc}$ film reflections.

The second key observation from Fig. 2 is that the $(0\ 1\ 3)_{pc}$ film peak has shifted upward and the $(0\ \text{-}1\ 3)_{pc}$ film peak has shifted downward with respect to the $(\pm 1\ 0\ 3)_{pc}$ film peaks. The $Q_\perp$ position of the latter is indicated by a dashed line for clarity. This behavior is consistent with in-plane crystallographic symmetry breaking, such as an orthorhombic distortion in case of $SrRuO_3$ [18]. The upwards shift of the $(0\ 1\ 3)_{pc}$ peak and downwards shift of the $(0\ \text{-}1\ 3)_{pc}$ peak with respect to the $(\pm1\ 0\ 3)_{pc}$ peaks imply that $(010)_{pc}$ planes have tilted towards $[0\ \text{-}1\ 0]_c$ with respect to $(010)_c$ ('c' indicates the substrate's cubic coordinates) by an angle $\delta$ [19], resulting in $\alpha_{pc} = 90° + \delta$ which is the angle between $[010]_{pc}$ and $[001]_{pc}$ (Fig. 3a). This in-plane symmetry breaking of the Ru-LSMO film is therefore ascribed to a monoclinic distortion, consistent with previous observations for LSMO [20,21].

A third observation from Figure 2 is the emergence of satellite film peaks along specific directions. The satellites are distinct around the $(\pm1\ 0\ 3)_{pc}$ film peaks, whereas no satellites are observed around the $(0\ \pm1\ 3)_{pc}$ peaks. This RSM feature is explained by a periodic structural modulation [10,20,21], and its exclusive appearance along the $[100]_{pc}$ direction indicates the existance of a 1-dimensional (1D) periodic structural domain array along or near this axis (Fig. 3b).

Altogether, the RSM analysis of the 48 nm 10% Ru-LSMO film points to a comressively-strained, coherent monoclinic (distorted orthorhombic) crystal structure (Fig. 3a) with a 1D crystallographic domain structure along $[100]_{pc}$. In monoclinic notation (subscript *m*), $[110]_m$ is parallel to $[001]_{pc}$, $[1\ \text{-}1\ 0]_m$ is parallel to $[010]_{pc}$, and $[100]_{pc}$ is parallel to $[001]_m$. In the interest of simplicity, we will continue with the 'pc' notation. These microstructural details play a key role in the magnetic anisotropy, to be discussed in the next section.

During the coherent film growth, the biaxial compressive strain applied to the film by the substrate compresses the Ru-LSMO pseudocubic unit cells along $[100]_{pc}$ and $[010]_{pc}$ and hence expands the pseudocubic unit cells along $[001]_{pc}$. This leads to the distorttion of the latttice by tilting and rotating of the $MnO_6$ and $RuO_6$ octahedra, resulting in monoclinic unit cells (Fig. 3a), with the monoclinic angle $\gamma_m$ being less than 90°. This interpretation agrees well with previous obsevations of simlilar lattice distortions and monoclinic unit cell formation in compressively strained LSMO films on LSAT substrates [20,21]. This kind of crystallographic anisotropy affects the spin orbit coupling in perovskite oxides, affecting the magnetic properties diffrerently along various crystallographic directions [10,21–24].

In addition to its lattice parameter mismatch with LSAT, the rhombohedral (bulk) LSMO unit cell further features a lattice *angle* mismatch with the cubic LSAT substrate. The lattice parameter mismatch induces biaxial compressives strain on the Ru-LSMO unit cells, whereas the lattice angle mismatch induces shear strain. The angle $\gamma_m$ becomes less than 90° to accommodate the lattice parameter mismatch. The monoclinic unit cells of Ru-LSMO can release a small amount of shear strain along $[010]_{pc}/[1\ \text{-}1\ 0]_m$ by changing the angle $\gamma_m$, resulting in an octahedral tilt. However, shear strain accumulates as the thickness of a Ru-LSMO film increases. To release this shear strain, periodic structural lattice modulation of the film occurrs along the lattice direction $[100]_{pc}$, at the cost of deviation of the angle ($\beta_{pc}$) between $[001]_{pc}$ and $[100]_{pc}$ from 90°. This occurs while keeping the $(100)_{pc}$ planes perpendicular to the substrate's surface plane $(001)_c$ (Fig. 3b) [10,20,21], resulting in satellite peaks in the Ru-LSMO films (Fig. 3). We note that the variation of the angle $\beta_{pc}$ is the key reason behind the broadening of the satellites (see Fig. S6 and the discussion therein). The lattice modulation along $[100]_{pc}$ induces periodic shifting of the centers of pseudocubic unit cells with periodicity $\tau$ along the lattice direction $[001]_{pc}$ (Fig. 3b). The separation between the main peak and satellite peaks of 48 nm 10% Ru-LSMO film in reciprocal space is $\Delta Q_{\parallel} = 0.0045 Å^{-1}$ (red arrows in Fig. 3), yielding a 1D structural modulation period of $\tau = (\Delta Q_{\parallel})^{-1} = 22$ nm $\pm$ 6 nm in real space (accounting for satellite broadening, see Fig. S6 and discussion therein).

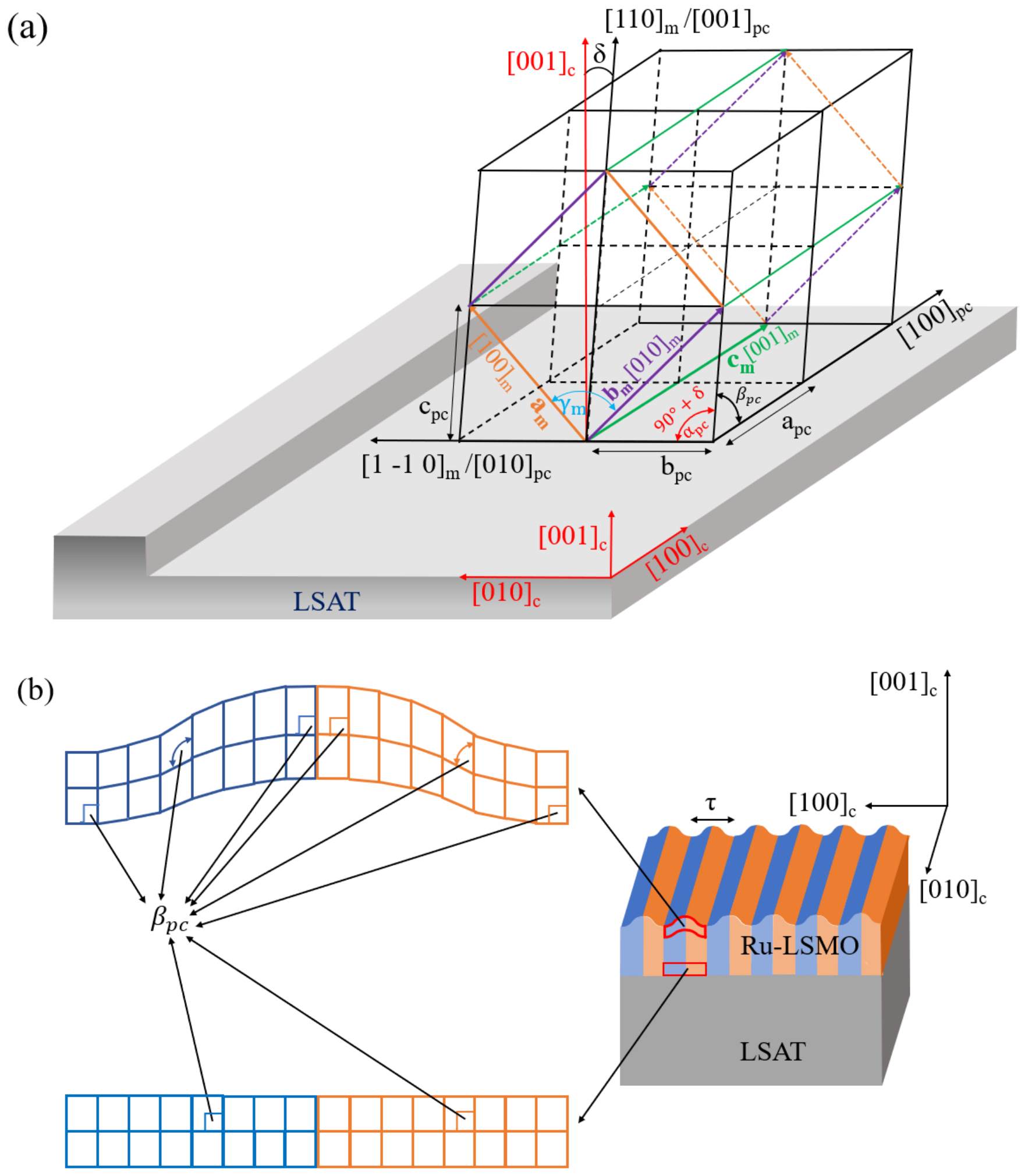

**Figure 3.** Schematic microstructure of Ru-LSMO coherently grown on LSAT. (a) Schematic of a (magnified) monoclinic unit cell of Ru-LSMO on a vicinal LSAT substrate. (b) 1D periodic structural modulation of Ru-LSMO films on a vicinal LSAT substrate. The substrate has miscut angle 0.1° with step edge direction being 8.0° clockwise from the lattice direction $[010]_c$.

The Ru-LSMO $(\pm 1\ 0\ 3)_{pc}$ main Bragg peaks are much more intense than their satellites, indicating that some volume of the film does not undergo the periodic lattice modulation. This is explained by the structural modulation starting above a critical thickness, releasing the (thickness-dependent) elastic energy. Indeed, an RSM analysis of 10.5 nm 10% Ru-LSMO film does not show any satellite features (Fig. S7) while retaining the monoclinic crystal structure. In addition, similarly to the 48 nm film, the 10.5 nm film has strong perpendicular magnetization component (Fig. S2, S1b), but it exhibits only weakly anisotropic in-plane magnetization. This suggests that the periodic structural domains are not necessary for the strong perpendicular component of magnetization, but they do play a role in the anisotropic in-plane magnetization, to be discussed later. The emergence of such lattice modulation, only above a critical thickness, has been reported in LSMO films [21,25,26], in good agreement with our observation. 1D periodic structural modulation has also been reported in $LaCoO_3$, showing such modulation in the entire film thickness [10].

Substrate miscut can play a significant role in the manifestation and orientation of such structural modulations in complex oxide films. To determine the miscut angle and the miscut direction, rocking curve measurements were performed on the LSAT substrate Bragg peaks (Fig. S8). The calculated miscut angle of the substrate is 0.1°, and the step direction is 8°±5° clockwise with respect to $[010]_c$. Therefore, the 1D periodic structural modulation occurs along the terraces, and the structural domains are perpendicular to the terraces. This picture is consistent with step-edge nucleation during the initial stage of the film growth [21].

The kind of structural modulation we observe in the Ru-LSMO films at room temperature has been observed in LSMO films on STO substrates as well [25]. In the case of STO substrates, the structural modulation disappears at low temperatures due to the structural phase transition (~105 K) and phonon softening in STO [25,27]. However, unlike the case of LSMO on STO, the pattern of structural modulation of LSMO films on LSAT susbtrates does not change with the temperature [25]. Also, the possibility of minute structural variations of LSAT substrates at low temperatures [28] has no expression in the M-T behavior (Figs. 1, S1 and S2a), thus validating the room temperature structural features for low temperatures as well.

## III.C. MICROSTRUCTURAL ORIGINS OF THE MAGNETIC ANISOTROPY

Having characterized the microstructure of the Ru-LSMO film, we now describe the microstructural mechanisms of its magnetic properties. The monoclinic (distorted orthorhombic) crystal structure reported here hosts the Glazer octahedral tilt system $a^+a^-c^-$, similarly to compressively strained LSMO and SRO films [20,22,29]. Therefore, we begin by considering two related magneto-crystalline anisotropy archetypes that host the same octahedral tilt system ($a^+a^-c^-$) as the present case: $E_{[001]pc} > E_{[010]pc} > E_{[100]pc}$ in LSMO films [11,21,30] versus $E_{[001]pc} < E_{[010]pc} < E_{[100]pc}$ in SRO films [22], when both are under compressive strain and the monoclinic lattice direction $[110]_m$ is along $[001]_{pc}$. As shown in Fig. 1, the anisotropy energy in the present case is $E_{[001]pc} < E_{[010]pc} < E_{[100]pc}$, which suggests that the present Ru-LSMO system behaves more closely to SRO than to LSMO, but with the distinct practical advantage of the much higher Curie temperature of LSMO. We will describe the atomic mechanisms of these archetypes, and from them we propose a mechanism for the presently observed TMA in Ru-LSMO.

On one hand, LSMO films with the $a^+a^-c^-$ tilt system were shown to induce weakly anisotropic in-plane magnetization, with $[100]_{pc}$ being magnetically easier than $[010]_{pc}$ [21]. In contrast, SRO films with the same tilt system exhibit TMA with strong perpendicular magnetization and anisotropic in-plane magnetization, with $[010]_{pc}$ being magnetically easier than $[100]_{pc}$ [22]. This comparison therefore suggests that the single ion anisotropy in the Ru ions is induced by compressive strain and it plays a key role behind the strong perpendicular magnetization in the 10% Ru-LSMO films [7]. This in turn implies that compressive strain, together with strong spin orbit coupling (SOC) of Ru ions, induces a preferred orientation in the Ru spins (to be discussed later), which then governs the orientation of Mn spins. The octahedral rotation (discussed in the next paragraph) which influences the Ru-Mn and Mn-Mn interactions further plays an important role behind the TMA in Ru-LSMO films.

While in-phase octahedral rotations enhance $e_g$-$e_g$ orbital overlap, the out-of-phase rotations enhance six out of nine $t_{2g}$-$t_{2g}$ orbital overlaps [21,22]. The Mn-Mn magnetic interaction in LSMO is based on the overlap of the $e_g$ orbitals, whereas the Ru-Ru magnetic interaction in SRO is based on the overlap of the $t_{2g}$ orbitals. The octahedral rotation $c^-$ about the $[001]_{pc}$ is out-of-phase in both LSMO and SRO films, but strong perpendicular magnetization is observed only in SRO. The in-plane magnetic easier axis of LSMO films is the axis ($[100]_{pc}$) around which the octahedral rotation ($a^+$) is in-phase, whereas the in-plane magnetic easier axis of SRO films is the axis $[010]_{pc}$ around which the octahedral rotation ($a^-$) is out-of-phase. The octahedral rotation induced orbital anisotropy, together with spin-orbit coupling (SOC), induce two opposite orders of magneto-crystalline anisotropy energy in these examples: $E_{[001]pc} > E_{[010]pc} > E_{[100]pc}$ in LSMO films versus $E_{[001]pc} < E_{[010]pc} < E_{[100]pc}$ in SRO films (both compressively strained). This difference is rooted in the different dominant orbitals (and their overlap): $e_g$ in LSMO versus $t_{2g}$ in SRO.

Here the 10% Ru-LSMO films (Fig. 1, Fig S2) exhibit the same order of magneto-crystalline anisotropy energy as observed in SRO films, opposite to the LSMO case. This suggests that the orbital anisotropy of the Ru 4d $t_{2g}$ orbitals and their interaction with Mn 3d $t_{2g}$ orbitals play a crucial role in the magnetic behavior of Ru-LSMO films. Indeed, it has recently been suggested that an antiferromagnetic interaction between Ru and Mn ions via $t_{2g}$ orbital overlap governs the magnetic properties of Ru-LSMO films [17]. We therefore propose that the monoclinic crystal structure together with the octahedral tilt system $a^+a^-c^-$ create the playground for Ru-Mn $t_{2g}$ interactions in 10% Ru-LSMO films, where the single ion anisotropy of the Ru ions dictates the spin orientation of Mn ions and determines the order of magneto-crystalline anisotropy energy here as $E_{[001]pc} < E_{[010]pc} < E_{[100]pc}$.

The Ru ion is driving the magnetic anisotropy by playing a dual role. First, the substitution of Mn by Ru increases the compressive strain which induces $t_{2g}$ orbital anisotropy by driving the monoclinic $a^+a^-c^-$ tilt structure, resulting in both in-plane and out-of-plane orbital anisotropy. The Ru 4d orbitals have an order of magnitude stronger SOC compared to the Mn 3d orbitals [31,32]. When strained, the $RuO_6$ octahedra are expected to translate their local orbital preference to the Ru spins via strong SOC of Ru ions, more effectively than the $MnO_6$ octahedra with the weaker SOC of Mn ions. Second, the Ru spins dictates the Mn spins via Ru-Mn $t_{2g}$ interactions [17]. The spatial distribution of the Ru 4d orbitals is wider than that of the Mn 3d orbitals, making Ru-Mn interactions stronger than Mn-Mn interactions (both of which occur via the oxygen anion) in Ru-LSMO. Moreover, the additional compressive strain induced by Ru substitution further increases the Ru-Mn interaction. Overall, the strain induced single ion anisotropy in Ru ions determines the magnetic anisotropy in 10% Ru-LSMO films via Ru-Mn $t_{2g}$ interactions.

Having discussed the role of Ru in magnetic anisotropy, we now consider the 1D periodic structural modulation (Fig. 3b), and its role in the anisotropic in-plane magnetization. This structural modulation appears in the thick Ru-LSMO film where the anisotropy of in-plane magnetization is relatively strong (Fig. 1), compared to the thin Ru-LSMO film (Figs. S2, S1b) where the 1D structural modulation is absent (Fig. S7). The weaker anisotropy of the in-plane magnetization is therefore ascribed to the monoclinic structure which exists in both films, as discussed earlier. The stronger anisotropy of in-plane magnetization in the thick film therefore shows correlation with the existence of the structural modulation. We ascribe the increased anisotropy of in-plane magnetization in the thick Ru-LSMO film to the additional out-of-phase octahedral rotation $a^-$ around the $[010]_{pc}$ axis as a result of the periodic variations in the angle $\beta_{pc}$ (Fig. 3b).

From the above discussion, we note that Ru plays an important role in the manifestation of TMA with strong perpendicular magnetization in Ru-LSMO. However, this raises a question whether the TMA observed here can be explained purely by strain. TMA with strong perpendicular magnetization in LSMO can be achieved without Ru substitution, albeit with high compressive strain (-2.2%) using $LaAlO_3$ (LAO) substrates [33]. However, in the present case, the TMA with strong perpendicular magnetization is observed under a moderate compressive strain (-0.41% [17]) with LSAT substrates. This comparison illustrates that strain alone cannot account for the observed TMA, highlighting the importance of Ru in translating the orbital anisotropy into magnetic anisotropy through strong SOC and more spread-out 4d orbitals. Ru substitution significantly enhances the strain induced perpendicular magnetization. From a practical perspective, high strain is less desirable as it limits the growth, thickness, and processing parameter space.

We now briefly highlight a possible technological implementation of the observed TMA in the 10% Ru-LSMO films. Field-free perpendicular magnetization switching through SOT holds promise for future low-power non-volatile magnetic memories. Practical implementation of such devices requires TMA with strong perpendicular magnetization component, which is usually achieved in ultrathin ferromagnetic metals (such as Co) or alloys (such as CoFeB) by complex geometries and structures [34–37], hindering their practical application. Therefore, materials that have tilted magnetic anisotropy (TMA) with a strong perpendicular magnetization component are of considerable advantage for such devices. For example, strong perpendicular magnetization of SRO [22,38] was recently utilized to demonstrate deterministic perpendicular magnetization switching through SOT in an all-oxide heterostructure at 70K [14]. However, the low Curie temperature ($T_C$) of SRO [14,22] is a major hurdle towards practical realization. Nakamura et al. showed that 10% Ru substitution in the high-$T_C$ material LSMO supports strong perpendicular magnetization up to much higher temperatures, but the in-plane magnetization was not addressed. Along with a strong perpendicular magnetization, anisotropic in-plane magnetization is crucial for deterministic perpendicular magnetization switching through SOT [13,15]. Moreover, the Curie temperature of the manganites can be engineered above the room temperature [39]. Therefore, the strong perpendicular magnetization along with the anisotropic in-plane magnetization in 10% Ru-LSMO could be utilized to fabricate the SOT switching devices, which could work much closer to (and potentially above) room temperature, paving the way towards practical applications.

## IV. SUMMARY AND CONCLUSIONS

We report TMA with strong perpendicular magnetization and anisotropic in-pane magnetization in 10% Ru-LSMO films under moderate compressive strain. The microstructure of the Ru-LSMO films was analyzed and correlated with their magnetic properties. We show how Ru magnifies the impact of strain, explaining the possible microstructural origin of magnetic anisotropy. We further illustrate how shear strain relaxation occurs above a certain thickness via the formation of 1D periodic structural modulation, which in turn plays a prominent role in the manifestation of anisotropic in-plane magnetization. Demonstrating and understanding the microstructural origin of TMA with strong perpendicular magnetization and anisotropic in-plane magnetization in 10% Ru-LSMO paves the way towards the realization of the practical oxide-based room temperature spintronic memories.

## ACKNOWLEDGEMENTS

This work was funded by the German Israeli Foundation (GIF Grant No. I-1510-303.10/2019). The authors thank Dr. Ionela Lindfors-Vrejoiu for growing the films used here and for fruitful discussions. We further thank Dr. Maria Koifman Khristosov and Dr. Anna Eyal for assistance with XRD measurements and magnetometry, respectively.

## References

[1] A. J. Millis, B. I. Shraiman, and R. Mueller, *Dynamic Jahn-Teller Effect and Colossal Magnetoresistance in $La_{1-x}Sr_xMnO_3$*, Phys. Rev. Lett. **77**, 175 (1996).

[2] M. Bowen, M. Bibes, A. Barthélémy, J.-P. Contour, A. Anane, Y. Lemaître, and A. Fert, *Nearly Total Spin Polarization in $La_{2/3}Sr_{1/3}MnO_3$ from Tunneling Experiments*, Appl. Phys. Lett. **82**, 233 (2003).

[3] J. H. Park, E. Vescovo, H. J. Kim, C. Kwon, R. Ramesh, and T. Venkatesan, *Direct Evidence for a Half-Metallic Ferromagnet*, Nature **392**, 794 (1998).

[4] A. Urushibara, Y. Moritomo, T. Arima, A. Asamitsu, G. Kido, and Y. Tokura, *Insulator-Metal Transition and Giant Magnetoresistance in $La_{1-x}Sr_xMnO_3$*, Phys. Rev. B **51**, 14103 (1995).

[5] Y. Tokura, *Critical Features of Colossal Magnetoresistive Manganites*, Reports Prog. Phys. **69**, 797 (2006).

[6] J. Dho, N. H. Hur, I. S. Kim, and Y. K. Park, *Oxygen Pressure and Thickness Dependent Lattice Strain in $La_{0.7}Sr_{0.3}MnO_3$ Films*, J. Appl. Phys. **94**, 7670 (2003).

[7] M. Nakamura, D. Morikawa, X. Yu, F. Kagawa, T. Arima, Y. Tokura, and M. Kawasaki, *Emergence of Topological Hall Effect in Half-Metallic Manganite Thin Films by Tuning Perpendicular Magnetic Anisotropy*, J. Phys. Soc. Japan **87**, 74704 (2018).

[8] M. Mathews, F. M. Postma, J. C. Lodder, R. Jansen, G. Rijnders, and D. H. A. Blank, *Step-Induced Uniaxial Magnetic Anisotropy of $La_{0.67}Sr_{0.33}MnO_3$ Thin Films*, Appl. Phys. Lett. **87**, 242507 (2005).

[9] B. Paudel, B. Zhang, Y. Sharma, K. T. Kang, H. Nakotte, H. Wang, and A. Chen, *Anisotropic Domains and Antiferrodistortive-Transition Controlled Magnetization in Epitaxial Manganite Films on Vicinal $SrTiO_3$ Substrates*, Appl. Phys. Lett. **117**, 81903 (2020).

[10] E.-J. Guo et al., *Nanoscale Ferroelastic Twins Formed in Strained $LaCoO_3$ Films*, Sci. Adv. **5**, eaav5050 (2022).

[11] F. Tsui, M. C. Smoak, T. K. Nath, and C. B. Eom, *Strain-Dependent Magnetic Phase Diagram of Epitaxial $La_{0.67}Sr_{0.33}MnO_3$ Thin Films*, Appl. Phys. Lett. **76**, 2421 (2000).

[12] M. Konoto, H. Yamada, K. Koike, H. Akoh, M. Kawasaki, and Y. Tokura, *Magnetic Quasidomain Structures in Ru-Doped $La_{0.6}Sr_{0.4}MnO_3$ Thin Films*, Appl. Phys. Lett. **93**, 252503 (2008).

[13] F. Wang, X. Zhang, Z. Zhang, and Y. Liu, *Deterministic Magnetization Switching by Spin–Orbit Torque in a Ferromagnet with Tilted Magnetic Anisotropy: A Macrospin Modeling*, J. Magn. Magn. Mater. **527**, 167757 (2021).

[14] L. Liu et al., *Current-Induced Magnetization Switching in All-Oxide Heterostructures*, Nat. Nanotechnol. **14**, 939 (2019).

[15] Z. Zheng et al., *Field-Free Spin-Orbit Torque-Induced Switching of Perpendicular Magnetization in a Ferrimagnetic Layer with a Vertical Composition Gradient*, Nat. Commun. **12**, 4555 (2021).

[16] O. I. Lebedev, J. Verbeeck, G. Van Tendeloo, C. Dubourdieu, M. Rosina, and P. Chaudouët, *Structure and Properties of Artificial* $[(La_{0.7}Sr_{0.3}MnO_3)_m(SrTiO_3)_n]_{15}$ *Superlattices on (001)SrTiO3.*, J. Appl. Phys. **94**, 7646 (2003).

[17] E. Hua et al., *Ru-Doping-Induced Spin Frustration and Enhancement of the Room-Temperature Anomalous Hall Effect in La2/3Sr1/3MnO3 Films*, Adv. Mater. **34**, 2206685 (2022).

[18] S. H. Chang, Y. J. Chang, S. Y. Jang, D. W. Jeong, C. U. Jung, Y. J. Kim, J. S. Chung, and T. W. Noh, *Thickness-Dependent Structural Phase Transition of Strained* $SrRuO_3$ *Ultrathin Films: The Role of Octahedral Tilt*, Phys. Rev. B - Condens. Matter Mater. Phys. **84**, 104101 (2011).

[19] S. W. Jin, G. Y. Gao, Z. Huang, Z. Z. Yin, X. Zheng, and W. Wu, *Shear-Strain-Induced Low Symmetry Phase and Domain Ordering in Epitaxial* $La_{0.7}Sr_{0.3}MnO_3$ *Thin Films*, Appl. Phys. Lett. **92**, 65 (2008).

[20] A. Vailionis, H. Boschker, W. Siemons, E. P. Houwman, D. H. A. Blank, G. Rijnders, and G. Koster, *Misfit Strain Accommodation in Epitaxial* $ABO_3$ *Perovskites: Lattice Rotations and Lattice Modulations*, Phys. Rev. B **83**, 64101 (2011).

[21] H. Boschker, M. Mathews, P. Brinks, E. Houwman, A. Vailionis, G. Koster, D. H. A. Blank, and G. Rijnders, *Uniaxial Contribution to the Magnetic Anisotropy of* $La_{0.67}Sr_{0.33}MnO_3$ *Thin Films Induced by Orthorhombic Crystal Structure*, J. Magn. Magn. Mater. **323**, 2632 (2011).

[22] W. Lu, W. Dong Song, K. He, J. Chai, C.-J. Sun, G.-M. Chow, and J.-S. Chen, *The Role of Octahedral Tilting in the Structural Phase Transition and Magnetic Anisotropy in* $SrRuO_3$ *Thin Film*, J. Appl. Phys. **113**, 63901 (2013).

[23] M. Ziese, I. Vrejoiu, and D. Hesse, *Structural Symmetry and Magnetocrystalline Anisotropy of* $SrRuO_3$ *Films on* $SrTiO_3$, Phys. Rev. B **81**, 184418 (2010).

[24] Q. Gan, R. A. Rao, C. B. Eom, L. Wu, and F. Tsui, *Lattice Distortion and Uniaxial Magnetic Anisotropy in Single Domain Epitaxial (110) Films of* $SrRuO_3$, J. Appl. Phys. **85**, 5297 (1999).

[25] F. Lan et al., *Observing a Previously Hidden Structural-Phase Transition Onset through Heteroepitaxial Cap Response*, Proc. Natl. Acad. Sci. **116**, 4141 (2019).

[26] O. I. Lebedev, G. Van Tendeloo, S. Amelinckx, F. Razavi, and H.-U. Habermeier, *Periodic Microtwinning as a Possible Mechanism for the Accommodation of the Epitaxial Film-Substrate Mismatch in the* $La_{1-x}Sr_xMnO_3/SrTiO_3$ *System*, Philos. Mag. A **81**, 797 (2001).

[27] U. Gebhardt, N. V Kasper, A. Vigliante, P. Wochner, H. Dosch, F. S. Razavi, and H.-U. Habermeier, *Formation and Thickness Evolution of Periodic Twin Domains in Manganite Films Grown on* $SrTiO_3$*(001)Substrates*, Phys. Rev. Lett. **98**, 96101 (2007).

[28] B. C. Chakoumakos, D. G. Schlom, M. Urbanik, and J. Luine, *Thermal Expansion of LaAlO3 and (La,Sr)(Al,Ta)O3, Substrate Materials for Superconducting Thin-Film Device Applications*, J. Appl. Phys. **83**, 1979 (1998).

[29] A. M. Glazer, *The Classification of Tilted Octahedra in Perovskites*, Acta Crystallogr. Sect. B Struct. Crystallogr. Cryst. Chem. **28**, 3384 (1972).

[30] S. K. Chaluvadi, F. Ajejas, P. Orgiani, S. Lebargy, A. Minj, S. Flament, J. Camarero, P. Perna, and L. Méchin, *Epitaxial Strain and Thickness Dependent Structural, Electrical and Magnetic Properties of La0.67Sr0.33MnO3 Films*, J. Phys. D. Appl. Phys. **53**, 375005 (2020).

[31] T. Mizokawa, L. H. Tjeng, G. A. Sawatzky, G. Ghiringhelli, O. Tjernberg, N. B. Brookes, H. Fukazawa, S. Nakatsuji, and Y. Maeno, *Spin-Orbit Coupling in the Mott Insulator ${\mathrm{Ca}}_{2}{\mathrm{RuO}}_{4}$*, Phys. Rev. Lett. **87**, 77202 (2001).

[32] T. Harano et al., *Role of Doped Ru in Coercivity-Enhanced $La_{0.6}Sr_{0.4}MnO_3$ Thin Film Studied by x-Ray Magnetic Circular Dichroism*, Appl. Phys. Lett. **102**, 222404 (2013).

[33] K. Steenbeck, T. Habisreuther, C. Dubourdieu, and J. P. Sénateur, *Magnetic Anisotropy of Ferromagnetic La0.7Sr0.3MnO3 Epitaxial Thin Films: Dependence on Temperature and Film Thickness*, Appl. Phys. Lett. **80**, 3361 (2002).

[34] G. Yu et al., *Switching of Perpendicular Magnetization by Spin–Orbit Torques in the Absence of External Magnetic Fields*, Nat. Nanotechnol. **9**, 548 (2014).

[35] L. You, O. Lee, D. Bhowmik, D. Labanowski, J. Hong, J. Bokor, and S. Salahuddin, *Switching of Perpendicularly Polarized Nanomagnets with Spin Orbit Torque without an External Magnetic Field by Engineering a Tilted Anisotropy*, Proc. Natl. Acad. Sci. **112**, 10310 (2015).

[36] S. Fukami, C. Zhang, S. DuttaGupta, A. Kurenkov, and H. Ohno, *Magnetization Switching by Spin–Orbit Torque in an Antiferromagnet–Ferromagnet Bilayer System*, Nat. Mater. **15**, 535 (2016).

[37] Y.-W. Oh et al., *Field-Free Switching of Perpendicular Magnetization through Spin–Orbit Torque in Antiferromagnet/Ferromagnet/Oxide Structures*, Nat. Nanotechnol. **11**, 878 (2016).

[38] S. Kunkemöller, F. Sauer, A. A. Nugroho, and M. Braden, *Magnetic Anisotropy of Large Floating-Zone-Grown Single-Crystals of $SrRuO_3$*, Cryst. Res. Technol. **51**, 299 (2016).

[39] A. Sadoc, B. Mercey, C. Simon, D. Grebille, W. Prellier, and M.-B. Lepetit, *Large Increase of the Curie Temperature by Orbital Ordering Control*, Phys. Rev. Lett. **104**, 46804 (2010).

# Supplementary information for

# Magnetic Anisotropy and Its Structural Origins in Ru-Substituted Manganite Films

Brajagopal Das,[1] Lena Wysocki,[2] Jörg Schöpf,[2] Lin Yang,[2] Daniel Jansen,[2] Amir Capua,[3] Paul H.M. van Loosdrecht,[2] Lior Kornblum[1]

[1]*Andrew & Erna Viterbi Department of Electrical & Computer Engineering, Technion Israel Institute of Technology, 3200003 Haifa, Israel*

[2]*University of Cologne, Institute of Physics II, 50937 Cologne, Germany*

[3]*Department of Applied Physics, The Hebrew University of Jerusalem, Jerusalem 91904, Israel*

## Temperature and magnetic field dependence of magnetization in 10% Ru-LSMO films

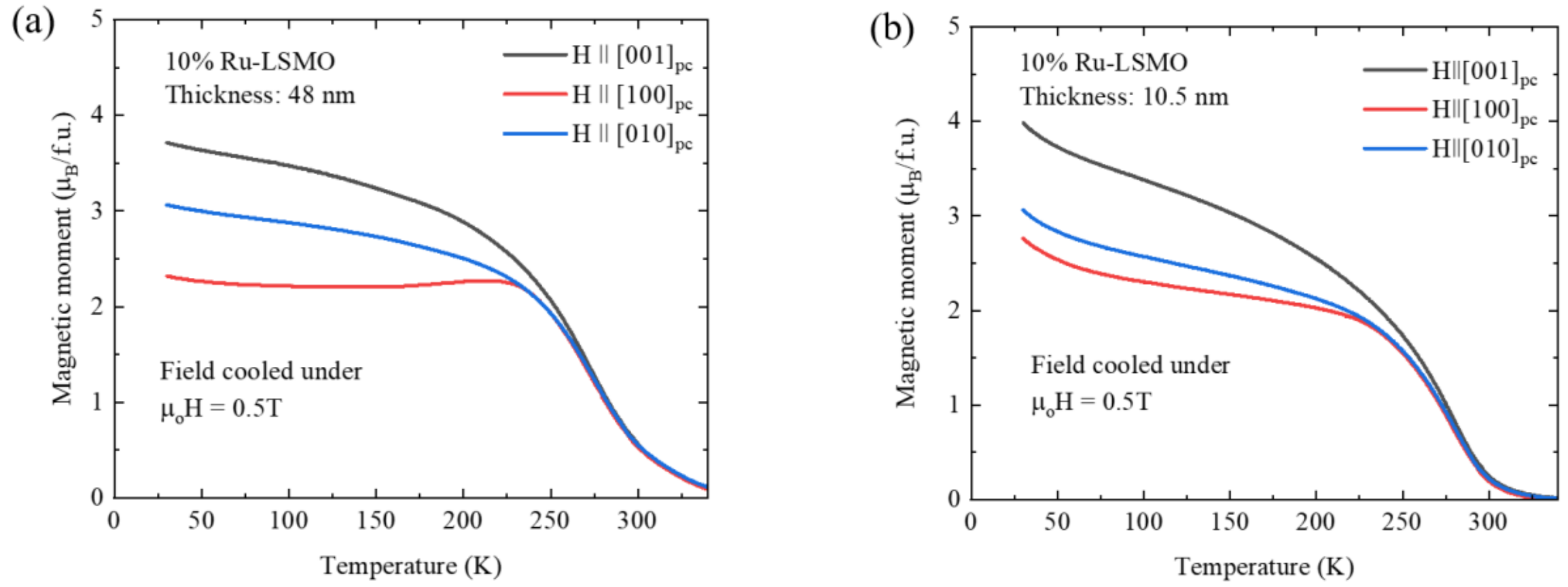

**Figure S1.** (a) Temeprature dependence of magnetization (M-T) in the thick (48 nm) 10% Ru-LSMO film and (b) thin (10.5 nm) 10% Ru-LSMO film. Before each M-T measurement, the samples were field cooled under 0.5 T and the measurements were performed during warm up under 0.5 T, which is well above the coercive fields of both the films. Therefore, these M-T curves reveal the magnetic anisotropy in both thick and thin 10% Ru-LSMO films. In the thin 10% Ru LSMO film, the anisotropic in-plane magnetization is originated from the monoclinic structural distortion, whereas the significant additional anisotropy of in-plane magnetization in the thick 10% Ru-LSMO film is originated from the 1D periodic structural modulation. The sudden upward trend at the low temperatures of M-T curves in the 10.5 nm 10% Ru-LSMO film is due to the impurity contribution from the LSAT substrate, which is illustrated in Fig. S3.

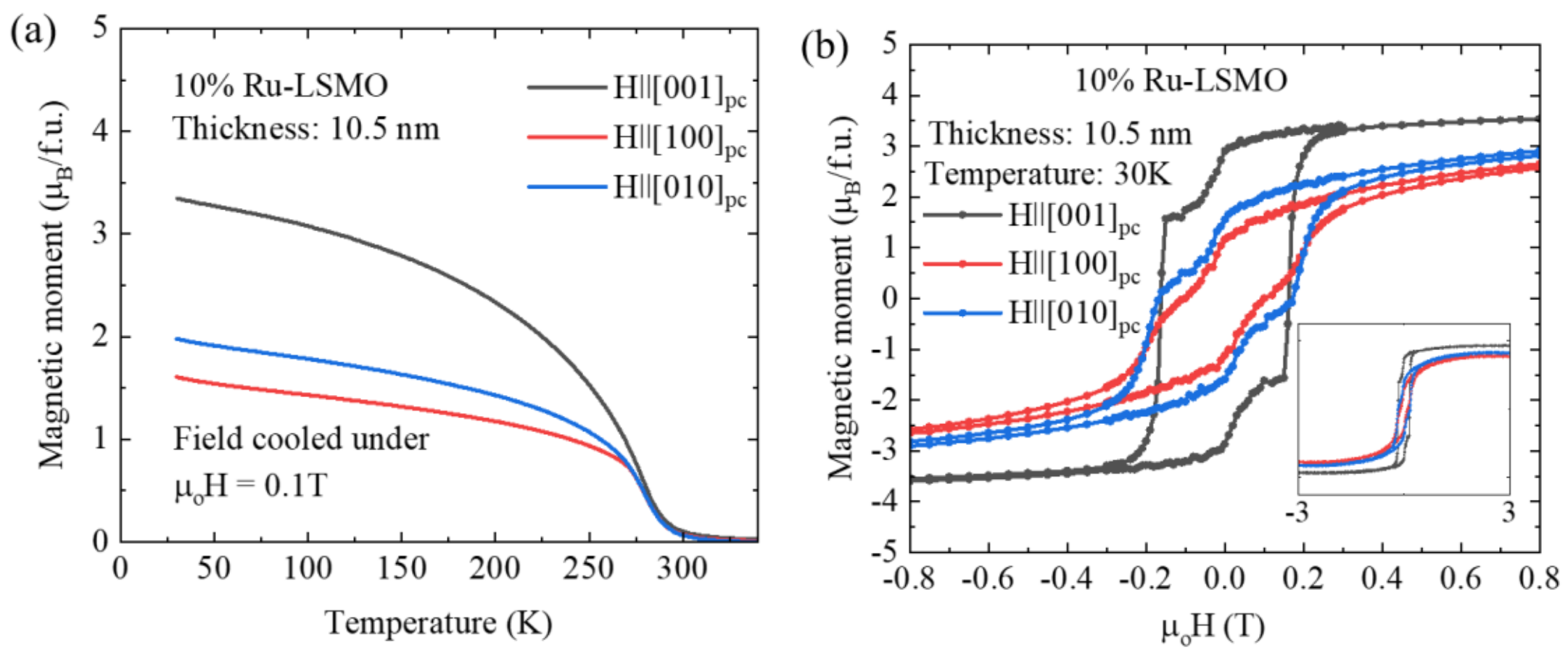

**Figure S2.** Magnetic properties of a thin 10% Ru-LSMO film. (a) Magnetization as a function of temperature (M-T) curves along the main pseudocubic lattice directions. (b) Magnetic field-dependent magnetization (M-H) loops at 30 K. Inset shows these M-H loops up to 3 T. The measurements sequence and details are identical to that of Fig. 1. The shoulder like features in the M-H loops was originated by impurity contribution from the substrate (see Kerr loop, Fig. S3).

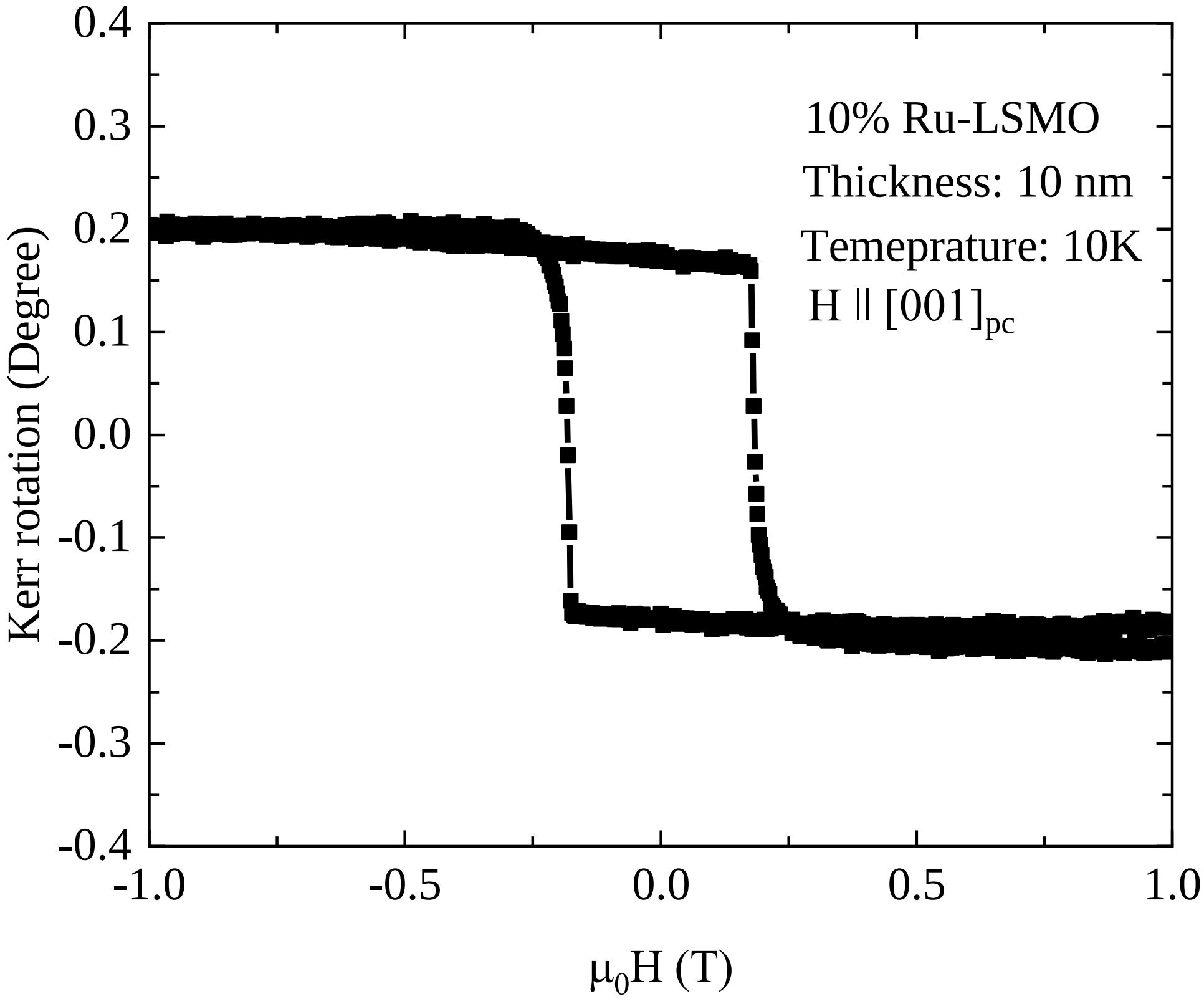

**Figure S3.** MOKE (Kerr rotation) loop of a 10 nm 10% Ru-LSMO film along $[001]_{pc}$; the penetration depth of light was around 50 nm (well above the film thickness of 10 nm). The MOKE loop shows no shoulder, suggesting that the shoulder observed in the M-H loops (Fig. S2b) of the 10.5 nm 10% film originated from the substrate impurity.

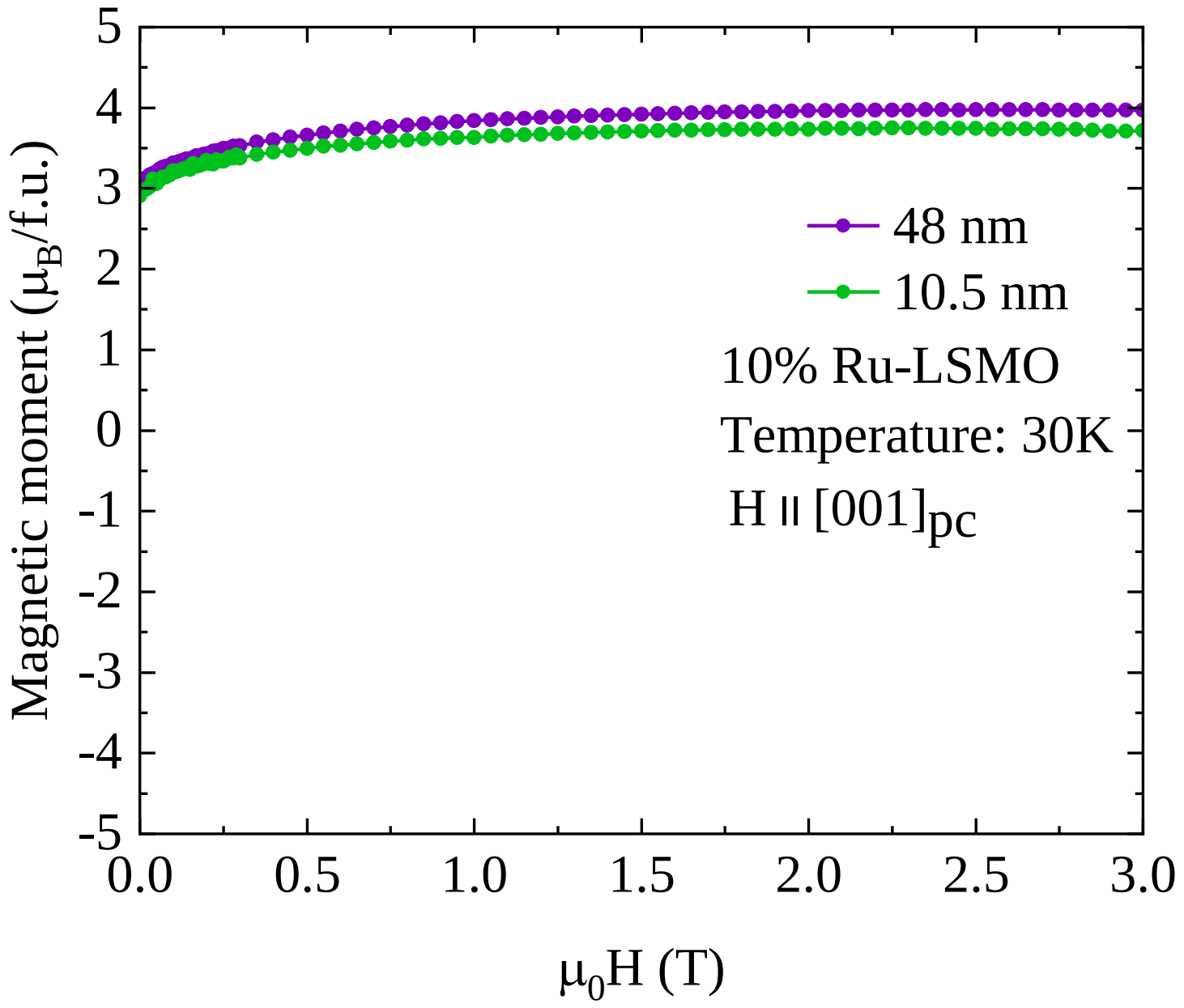

**Figure S4**. Comparison between the M-H curves of the thin and thick 10% Ru-LSMO films from saturation to the remnant magnetization along $[001]_{pc}$. Both the curves are pretty close to each other and almost coincide at the zero field, suggesting that the shape anisotropy doesn't have any significant impact on the perpendicular magnetization. Moreover, the shape anisotropy diminishes with increasing temperature [1], suggesting that the role of shape anisotropy can be ignored while describing the origin of observed magnetic anisotropy in our 10% Ru-LSMO films.

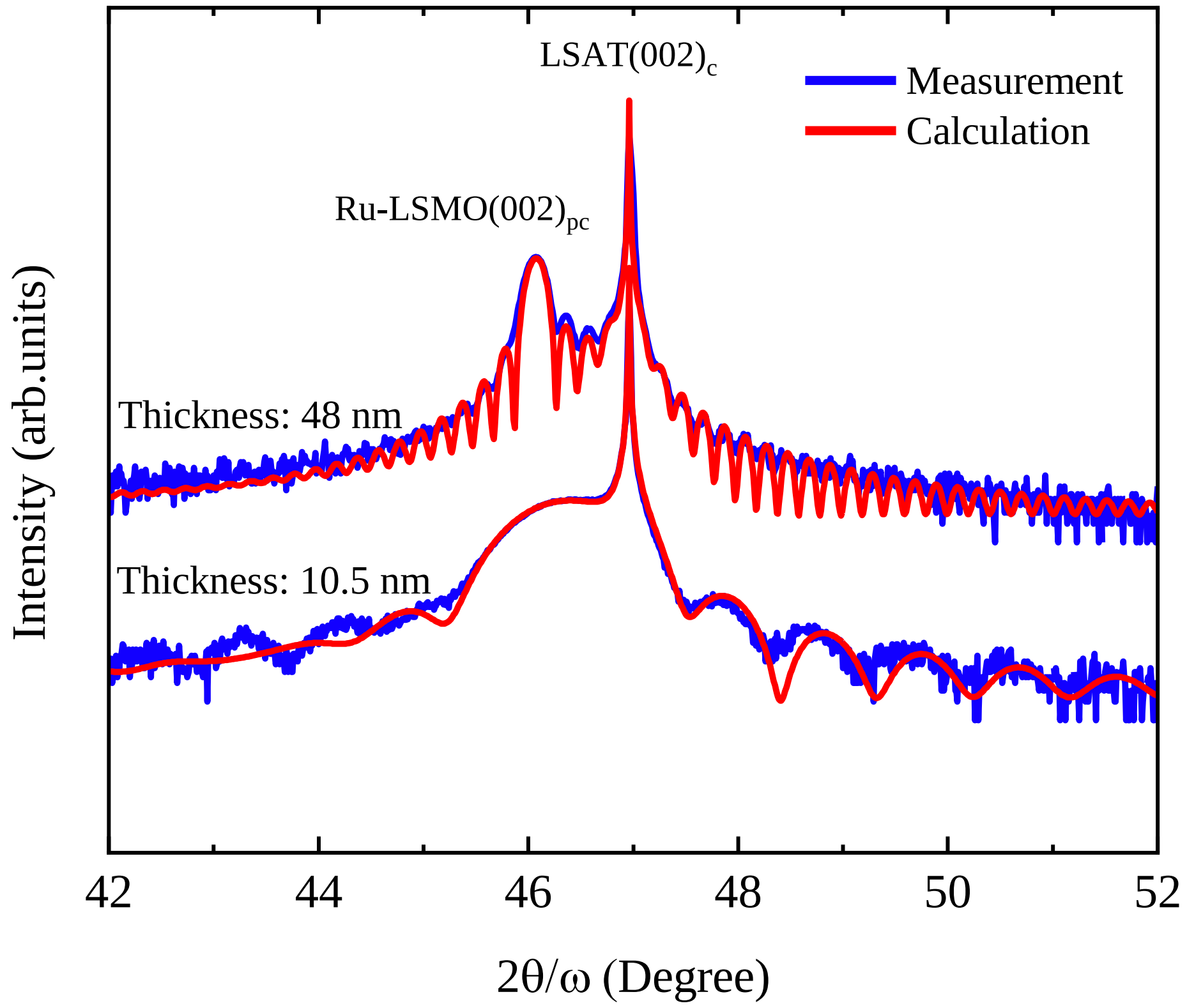

**Figure S5.** XRD (2θ/ω) scans of 10% Ru-LSMO films around their $(002)_{pc}$ reflections. The $(002)_{pc}$ peak of the thick and thin 10% Ru-LSMO films and the $(002)_{c}$ Bragg peak of the LSAT substrate are shown in the figure, respectively. Here the subscript 'pc' stands for pseudocubic lattice notation and 'c' for cubic. The sharp features and prevalence of Laue oscillations confirm the high structural quality of the epitaxial films.

## Satellite broadening

The following paragraph discusses the broadening of the satellites (compared to the film peak) in the RSM of Fig. 2.

The angle $\beta_{pc}$, estimated from the peak position of the satellite peaks, is 90.95° and 89.05° for the left and right satellite peaks respectively, whereas angle $\beta_{pc}$ is 90° for the main Bragg peak $(001)_{pc}$. The contribution of finite length to the broadening ($B_t$) of the XRD peaks can be estimated with the formula $\mathrm{FWHM} = \frac{0.885\lambda}{t\cos\theta_B}$, where λ is the x-ray wavelength, t the film thickness for the main Bragg peak and the structural domain periodicity for the satellites, and $\theta_B$ the Bragg angle [2]. Since structural modulation is in-plane, for the satellites, t is the periodicity of the satellites. The periodicity of the satellites is $22\ \mathrm{nm} \pm 6$; considering $t = 16, 22, and\ 28\ nm,$ the range of values for $B_t$ are $\sim$ 0.09°, 0.06°, and 0.05° respectively, which are much less than the observed broadening of the satellite peaks (~0.42°, Table S1). On the other nhand, the finite length broadening of the main Bragg peak is $\sim$ 0.03° (Fig. S6, Table S1). Therefore, 0.06° being the total broadening of the main Bragg peak, a rough upper bound for all other possible source of broadening including instrumental broadening is $\sim$ 0.03°, which can be cosidered to be same for the satellites [2]. This implies that the main source of satellite broadening is not due to the finite length broadening ($B_t$). Therefore, we consider variations in $\beta_{pc}$ as the main source of the satellite broadening in Fig. 2.

In addition, since the domain size (and its uncertainty) are much larger than the Ru-LSMO unit cell, it is not possible to determine whether they are fully commensurate, but the overall high structural quality (Fig. S5) suggests that it is likely the case.

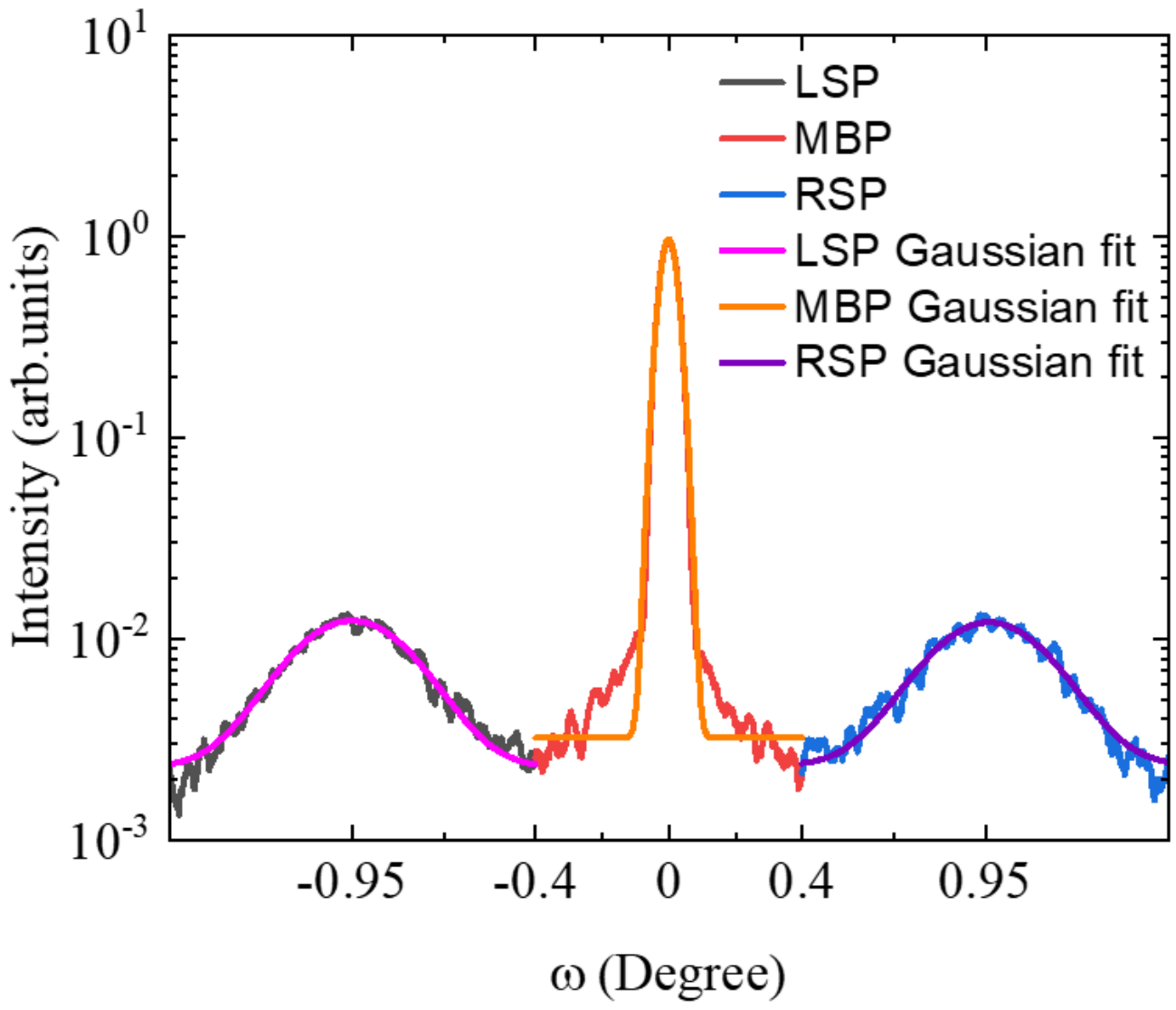

**Figure S6.** Rocking curve measurement of 48 nm 10% Ru-LSMO film around $(001)_{pc}$ peak. Gaussian fitting of the main Bragg peak (MBP), left satellite peak (LSP), and the right satellite peak (RSP) have been performed to find the FWHM of the peaks; the fitting parameters are tabulated below (Table S1).

**Table S1.** Individual Gaussian fitting of main Bragg peak $(001)_{pc}$, left satellite peak (LSP), and right satellite peak (RSP)

| | LSP | MBP | RSP |
|---|---|---|---|
| Peak position (θ) | 10.33° ± 0.01° | 11.28° | 12.23° ± 0.01° |
| Standard deviation (σ) | 0.18° | 0.027° | 0.18° |
| Full width at half maximum (FWHM) | 0.42° | 0.06° | 0.42° |

## RSM analysis of a thin 10% Ru-LSMO film

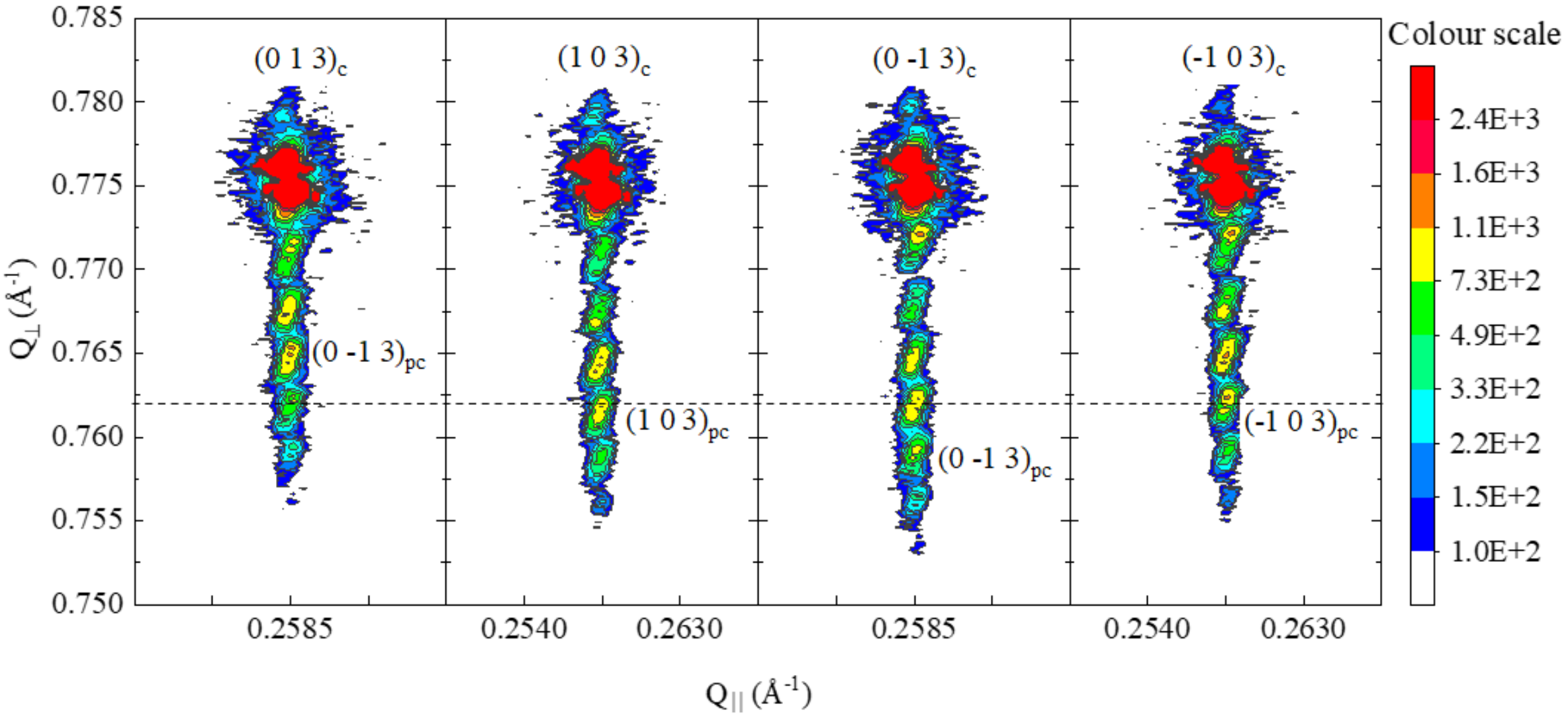

**Figure S7.** Off-specular RSMs of 10.5 nm 10% Ru-LSMO film around the $(0\ 1\ 3)_c$, $(1\ 0\ 3)_c$, $(0\ -1\ 3)_c$, and $(-1\ 0\ 3)_c$ reflections of LSAT. Intensities are presented on the logarithmic scale. No satellites are observed in the RSMs of the thin 10% Ru-LSMO film, indicating the absence of 1D periodic structural modulation, which is observed in the RSMs of thick 10% Ru-LSMO film (Fig. 2). Moreover, the $(013)_{pc}$ and $(0\ -1\ 3)_{pc}$ peaks have shifted upwards and downwards (respectively) in comparison to the $(103)_{pc}$, $(-1\ 0\ 3)_{pc}$ peaks (dashed line), indicating the monoclinic structure of thin 10% Ru-LSMO film [3–5].

## Measurement of the substrate miscut angle

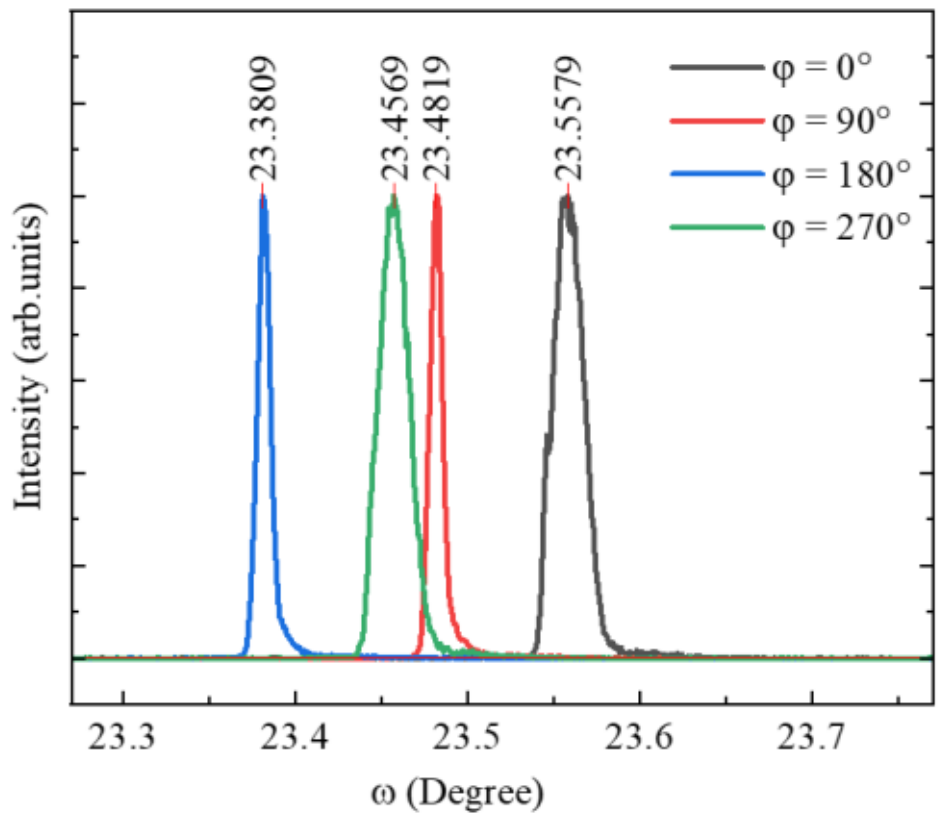

**Figure S8.** Rocking curve measurements (ω-scans) around the $(002)_c$ peak of the LSAT substrate in all four directions, indicating the presence of miscut on the substrate with 48 nm 10% Ru-LSMO film; the angle φ is measured clockwise with respect to ($[010]_c$). The estimated miscut angle is 0.1°, and the direction of terrace (step) is 8° clockwise with respect to the $[100]_c$ ($[010]_c$). The miscut angle and the step directions of all the substrates are listed in Table S2.

**Table S2.** The estimated miscut angles and directions of terraces of two substrates from the rocking curve measurements have been enlisted blow. The ±5° uncertainty is the estimated upper bound for the uncertainty in the positioning of the substrate on the diffractometer stage.

| Sample | Substrate | Miscut angle (Degree) | Direction of terraces with respect to $[100]_c$ |
|---|---|---|---|
| 48 nm 10% Ru- LSMO | LSAT(001) | 0.1 | 8° ± 5° (clockwise) |
| 10.5 nm 10% Ru- LSMO | LSAT(001) | 0.1 | 16° ± 5° (clockwise) |

## References:

[1] L. He and C. Chen, *Effect of Temperature-Dependent Shape Anisotropy on Coercivity for Aligned Stoner-Wohlfarth Soft Ferromagnets*, Phys. Rev. B - Condens. Matter Mater. Phys. **75**, 184424 (2007).

[2] A. J. Ying, C. E. Murray, and I. C. Noyan, *A Rigorous Comparison of X-Ray Diffraction Thickness Measurement Techniques Using Silicon-on-Insulator Thin Films*, J. Appl. Crystallogr. **42**, 401 (2009).

[3] E.-J. Guo et al., *Nanoscale Ferroelastic Twins Formed in Strained $LaCoO_3$ Films*, Sci. Adv. **5**, eaav5050 (2022).

[4] A. Vailionis, H. Boschker, W. Siemons, E. P. Houwman, D. H. A. Blank, G. Rijnders, and G. Koster, *Misfit Strain Accommodation in Epitaxial $ABO_3$ Perovskites: Lattice Rotations and Lattice Modulations*, Phys. Rev. B **83**, 64101 (2011).

[5] H. Boschker, M. Mathews, P. Brinks, E. Houwman, A. Vailionis, G. Koster, D. H. A. Blank, and G. Rijnders, *Uniaxial Contribution to the Magnetic Anisotropy of $La_{0.67}Sr_{0.33}MnO_3$ Thin Films Induced by Orthorhombic Crystal Structure*, J. Magn. Magn. Mater. **323**, 2632 (2011).